# Noctilucent Clouds Altitude and Particle Size Mapping Based on Spread Observations by Ground-based All-sky Cameras

Oleg S. Ugolnikov

Space Research Institute, Russian Academy of Sciences,
Profsoyuznaya st., 84/32, Moscow, 117997, Russia
E-mail: ougolnikov@gmail.com

**Abstract**

We suggest the procedure of building the maps of noctilucent clouds (NLC) zonal and meridional velocity, mean altitude and particle size based on three-color photometry by identical all-sky RGB-cameras separated by 115 km in a close-meridional direction. The procedure is applied to the bright NLC event on July 3, 2023. The altitude is measured by precise triangulation technique, and effective particle radius is estimated by comparison of each definite NLC fragment intensity and colors at different scattering angles as it is registered from different observation sites. The results are compared with existing photometric methods for average altitude and particle size measurements. A significant difference in evening and morning twilight NLC parameters is found, which is discussed in comparison with existing analysis of diurnal NLC variations.

**1. Introduction**

Noctilucent clouds (NLC) are the highest clouds in the atmosphere of the Earth. They were initially observed in late 19$^{th}$ century (Leslie, 1885). The clouds remained sunlit during the dark stage of twilight, which showed that they were higher than 80 km above the Earth's surface. Jesse (1896) was the first to measure NLC altitude using the triangulation technique. The clouds consist of water ice (Hervig et al., 2001), and temperatures below 145-150K are required for particles to freeze. There were no reliable mentions about NLC observations before 1880s (Dalin et al., 2012), their appearance and increasing occurrence rate can be related to climate changes in the entire atmosphere and positive trends of greenhouse gases (Roble and Dickinson, 1989).

Occurrence rate of NLC tended to increase during 20$^{th}$ century (Thomas and Olivero, 2001), but no visual trends were seen later (Romejko et al., 2003; Pertsev et al., 2014; Dalin et al., 2020). Ultraviolet measurements (SBUV, DeLand and Thomas, 2015) had shown an increase of total ice water content in the mesosphere, this was basically driven by greenhouse cooling and an increase of water vapor during the last decades (Hervig et al., 2016). These two processes lead to different effects on NLC characteristics (Lübken et al., 2018): increase of $CO_2$ causes the "atmospheric shrinking" and a negative trend of NLC mean altitude without a significant change of occurrence rate, while particle size and brightness are most sensitive to the water vapor density (von Zahn et al., 2004).

Altitude and particle radius have been measured for NLC for decades, and the trends are noticeable (Berger and Lübken, 2015). The most accurate method of altitude measurement is lidar sounding (von Cossart et al., 1997), particle size can also be defined if a multi-wavelength system is used (von Cossart et al., 1999; Alpers et al., 2000). Cross-polarization measurement mode allows studying the shape of ice particles (Baumgarten et al., 2002). Lidar analysis retrieves the local properties of NLC particles above the observer with fine time and altitude resolution.

NLC properties can be globally studied by spaceborn remote sensing. Multi-year data on occurrence rate and particle size was provided by the SCIAMACHY instrument onboard the Envisat satellite (von Savigny et al., 2004; Robert et al., 2009); total ice volume was measured by the MIPAS instrument (García-Comas et al., 2016); altitude and occurrence rate were fixed by the GOMOS instrument (Perot et al., 2010) of the same satellite. The longest data array was received in the AIM satellite mission (Russell et al., 2009) with the multi-angle instrument CIPS (Bailey et al., 2009; Lumpe at al., 2013) and the limb-occultation instrument SOFIE (Hervig et al., 2009, 2013). The satellite had finished working in early 2023. CIPS measurements can be used for building the maps of particle size, where radius can anti-correlate with albedo (Rusch et al., 2017; Gao et al., 2018). Changes in the relative position of Sun, observer, and cloud fragment make it possible to fix the phase dependence of scattering, which is important for particle size definition. The same technique was used in rocket UV-measurements (Gumbel et al., 2001; Gumbel and Witt, 2001; Hedin et al., 2008). Size estimation by comparison of scattering at different angles is also the basic principle of this work.

Triangulation method of altitude measurements has been used since the early years of NLC observations. However, good accuracy of altitude requires fine coordinate data of cloud elements. For example, if we have the observational base length equal to the NLC altitude (80-82 km), then we have to know the position of a single NLC element with an accuracy of 4 arcmin to find the altitude with an error of 0.2 km if the cloud is directly above the baseline. Error increases if the cloud is remote or if the baseline is shorter than NLC altitude. The characteristic angular size of NLC fragments is usually much larger than several arcmin, accurate field correlation analysis is necessary.

Altitude measurement of NLC was an aim of a number of spread cameras imaging (Dubietis et al., 2011; Dalin et al., 2013). Suzuki et al. (2016) used the correlation function method and fine-resolution cameras to find the mean value for the cloud ensemble of the first observed NLC in Japan with an accuracy of 0.1 km. If the cloud field is bright and well fragmented, we can run this procedure for each separate element and build the map of NLC altitude in the vicinity of observation sites. If we process the raw photometric data (not photo images) and know the spectral bands of sensors and camera optics, we can compare the brightness and color ratio of the same fragment from different sites and build the map of the effective particle radius. In this work we suggest the technique and apply it to the brightest noctilucent clouds of 2023 in the observation place.

## 2. Observations

Spread observations of a bright NLC field were conducted during the night of July 3, 2023. NLC brightness and color ratios were fixed by four identical Mi Sphere Cameras (the same devices as used by Ugolnikov, 2023ab) with a field size diameter about 190° and a resolution of 3456x3456 pixels. Two cameras were placed in the northern observation point (Site A, 56.61°N, 36.36°E, 135 m a.s.l.), and two other cameras operated in southern site (Site B, 55.58°N, 36.56°E, 190 m a.s.l.), the same as in (Ugolnikov, 2023 ab). The distance between the sites by the Earth's surface was equal to 114.7 km, or about 1.4 of the mean NLC altitude. This length and almost meridional direction of the observational baseline make this configuration optimal for such analysis.

The effective wavelengths of the B, G, and R channels for NLC case are equal to 474, 529 and 578 nm, respectively, however, integration over the spectral bands is processed during the analysis (Ugolnikov, 2023a). Nighttime images are used for star positions astrometry and photometry. The NLC triangulation procedure requires precise coordinate measurements. Updated 14-parameter model is used for conversion between frame and sky coordinates with a mean accuracy about 0.7 pixel (or 2 arcmin) for zenith angles up to 70°. It is based on the data of thousands star positions in the images of each camera during the night. Stellar photometry is used for the determination of atmospheric extinction in the B, G, and R channels in observation sites.

Measurements were held from evening twilight to morning twilight. Bright noctilucent clouds were observed throughout the night, but scattered tropospheric clouds restricted the spatial and temporal simultaneous coverage of NLC from two sites during the evening twilight. The skies were clear of tropospheric clouds in the morning. Exposure time varied from 1 second during the light twilight phase, when NLC became visible (solar zenith angles about 95°), to 4 seconds during the night. Imaging frequency varied between 4 and 5 frames per minute for each camera.

Average values of NLC "umbral" altitude and effective particle radius for the definite twilight can be found by NLC color analysis for different scattering angles and twilight stages as the Sun immerses into the shadow of the ozone layer and then dense troposphere (Ugolnikov, 2023a). The procedure uses EOS Aura\MLS data on temperature and ozone, Suomi NPP/OMPS data on stratospheric aerosol (NASA Goddard, 2023), and mean $NO_2$ profiles for definite time and location (Gruzdev and Elokhov, 2021). Evening data is spatially and temporally restricted by tropospheric clouds, however, averaging over the data of four cameras at two sites gives the altitude $83.55 \pm 0.10$ km and the particle radius $73.0 \pm 1.4$ nm. There is no restriction in the morning data, and we find the altitude $81.33 \pm 0.16$ km and the particle radius $84.9 \pm 5.4$ nm. Here we note that the morning altitude is close to the mean altitude obtained by this method for different NLC events in recent years, while the evening altitude is maximal over the same sample of observations. We see a significant decrease of NLC altitude during the night, this can be a part of the diurnal NLC variation observed in lidar experiments (see below). In this work altitudes of NLC will be found for cloud fragments separately by using the triangulation technique. An effective particle radius can also be defined for each fragment.

## 3. Triangulation procedure

Triangulation altitude of a definite NLC fragment is this work is found in a following form:

$$H = H_0 + \Delta H \qquad (1)$$

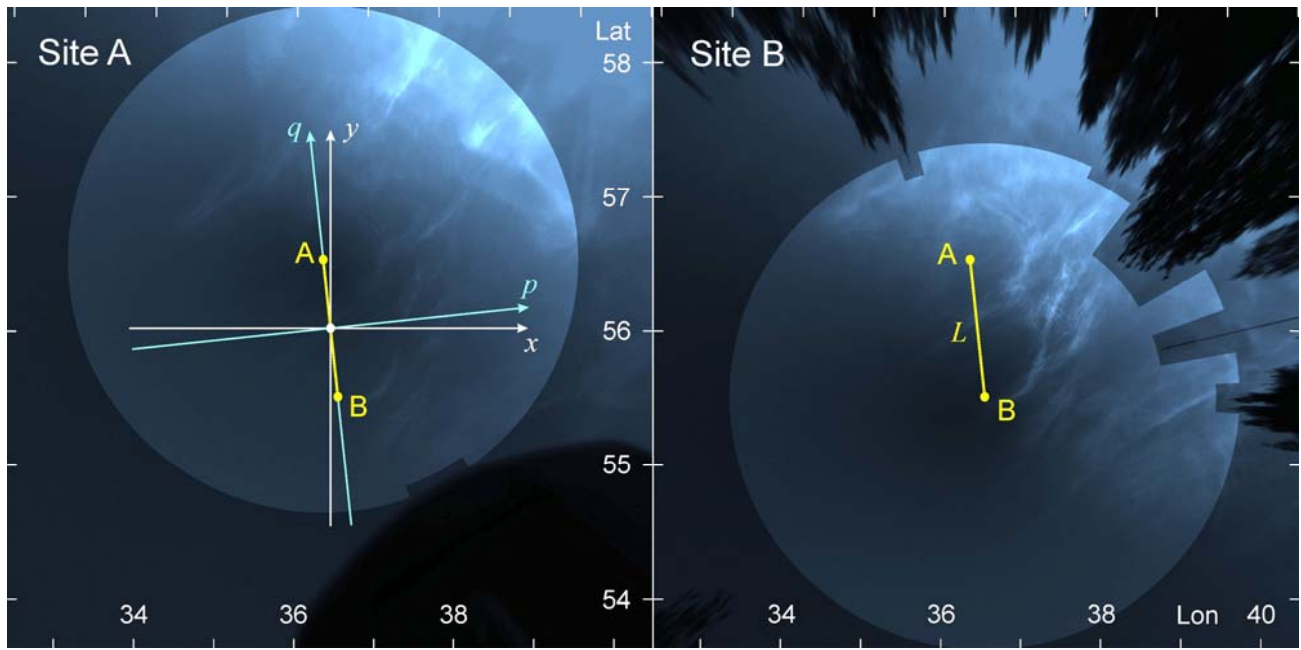

*Figure 1. Images of noctilucent clouds (July 3, 23:36 UT) from observation sites A and B projected onto the cloud layer with a priori altitude. The areas being processed are marked brightly. The coordinates used in this work are denoted in the left image.*

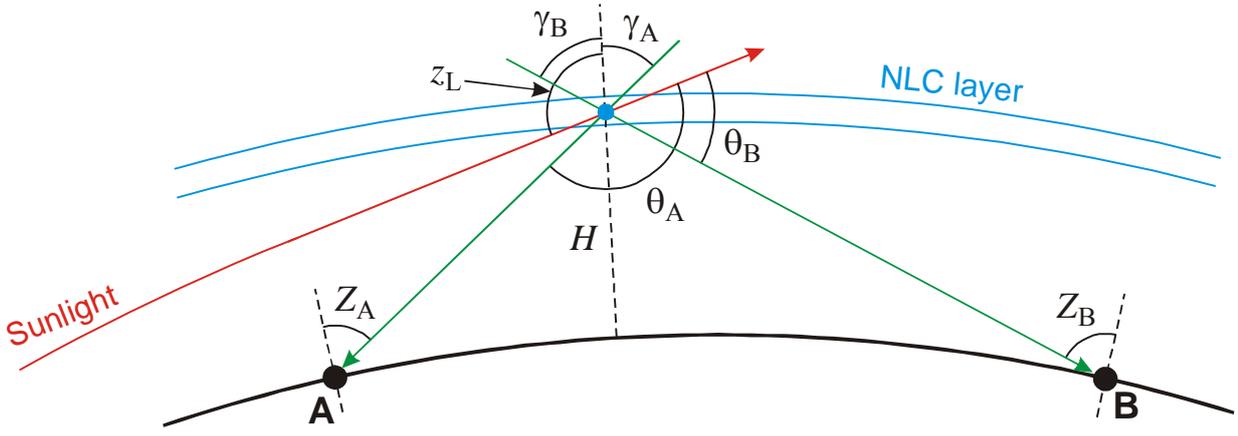

*Figure 2. The scheme of measurements for the case where a cloud fragment is above the baseline.*

Here $H_0$ is NLC altitude *a priori*, and the correction $\Delta H$ is fixed in a procedure. It works well if $\Delta H$ is about several kilometers and the assumption $\Delta H \ll H_0$ can be used. We chose the value $H_0$=81.33 km, found by umbral analysis during the tropospheric cloud-free morning twilight. It is close to the mean altitude of NLC measured during the past several years. As we will see below, it is also close to the average altitude of clouds during the night of July 3, 2023.

Sky image with NLC taken at a certain moment and location can be recalculated to the surface with the fixed altitude $H_0$ above sea level. We do it using the ellipsoidal model of the Earth and taking into account the small altitude of the observation points and the refraction of the light by the trajectory from NLC to the observer. Figure 1 shows the examples of these maps based on the pictures from two observational sites taken at the same moment in the morning twilight. The spatial resolution of this procedure is 0.08 arcmin by Earth meridian arc, which corresponds to a distance of about 150 meters in the NLC layer. If the altitude of all noctilucent clouds in a field was equal to $H_0$, the cloud patterns in both pictures would coincide. Actually, there is a shift depending on cloud location, this is the basis for the $\Delta H$ calculation.

Before correlation analysis of two observed fields is started, background subtraction and field calibration must be done. Background subtraction procedure for all-sky images was described in detail for polar stratospheric clouds analysis (Ugolnikov et al., 2021) and then used for NLC photometry (Ugolnikov, 2023ab). The basic idea of this procedure is the separation of NLC and background points along the definite line in the sky, followed by 6-degree polynomial expression of background in bands 1 (B), 2 (G), and 3 (R). Here we run the separation procedure along the horizontal lines $y$=const in Figure 1 for each definite $y$, and then take all background points $(x, y)$ to express the background as a polynomial by least squares method:

$$I_{0(1,2,3)} = \sum_{m+n \leq 6} W_{mn(1,2,3)} x^m y^n \tag{2}$$

To minimize the influence of large-scale NLC brightness gradients on the triangulation procedure, the special normalization is applied. Let the fragment of NLC is seen at zenith angle $Z_A$ ($Z_B$) from Site A (B). Figure 2 shows the picture for the simple case if the fragment is straight above the observational baseline. The intensity of the fragment (per solid angle unit) is

$$J_{A,B(1,2,3)} = I_{A,B(1,2,3)} - I_{0A,B(1,2,3)} =$$
$$= const \cdot K_{1,2,3}(z_L(x,y)) \cdot \frac{1}{\cos \gamma_{A,B}} \cdot \exp(-\tau_{A,B(1,2,3)} \cdot AM(Z_{A,B})) \cdot S_{1,2,3}(\theta_{A,B}, r) \tag{3}$$

Following Ugolnikov (2023a), $K(z_L)$ is the factor of illumination of the fragment by the Sun. It is defined by the local solar zenith distance $z_L$ visible from the cloud. This factor is the same for sites A and B, if we consider the definite moment of time. The second term $(1/\cos\gamma)$ reflects the position of optically thin NLC layer relative to the line of sight. We note that $\gamma_{A,B} \approx Z_{A,B}$ if the cloud is high above the horizon, however, this simplification is not used during the analysis. Here we assume the angle between the NLC layer and the Earth's surface to be small. Actually, the picture can be interfered by non-horizontal wave structure (Shevchuk et al., 2020), Kelvin-Helmholz instability (Baumgarten and Fritts, 2014) or possible bifurcations of the NLC layer (Gao et al., 2017). Due to this reason, the triangulation data is checked by correlation criterion (see below), and color data is used primarily to intensity data in the size definition procedure.

The following term $\exp(-\tau)$ in Equation (3) reflects the extinction of the NLC scattered emission in the lower atmosphere, optical depths $\tau$ are found by stellar photometry in nighttime images. $AM(Z)$ is the atmospheric mass for zenith angle $Z$, it is close to $(1/\cos Z)$, however, we do not use this flat simplification again. Finally, $S$ is the Mie scattering coefficient defined by the spectral interval (1, 2, or 3), scattering angle $\theta$, and effective particle radius $r$. We note that the surface brightness of the NLC fragment is measured, instrumental angular scale (pixel) is much smaller than the character size of the fragment. In this case, measured surface brightness does not directly depend on the distance from observation place to the cloud.

Analogously to the altitude $H_0$, we take *a priori* value of effective particle radius $r_0$ from morning data (85 nm, see above) and calculate the normalized surface brightness of NLC small element:

$$J_{A,B(1,2,3)N} = \frac{J_{A,B(1,2,3)} \cdot \cos\gamma_{A,B}}{S_{1,2,3}(\theta_{A,B}, r_0)} \cdot \exp(\tau_{A,B(1,2,3)} \cdot AM(Z_{A,B})). \tag{4}$$

If we had no instrumental errors and the NLC layer was horizontal with the altitude $H = H_0$ and fixed particle radius $r = r_0$, then the values of $J_{N(1,2,3)}$ as the functions of coordinates $(x, y)$ would be equal in sites A and B.

To find meridional and zonal NLC pattern velocities and then triangulation altitudes, the fragmentation procedure is used. We use the maps of total normalized brightness of NLC that is equal to the sum of normalized intensities in three spectral bands:

$$J_N(x, y) = J_{N1}(x, y) + J_{N2}(x, y) + J_{N3}(x, y) \tag{5}$$

Further triangulation analysis is related to the total normalized intensity, and we drop out an index "N". This field must be divided into the fragments, where the correlations will be calculated. The graphic scheme of this separation is shown in Figure 3. If we have normalized NLC intensity maximum $J$, we include in its fragment all adjacent maxima and areas with the intensity higher than $j = kJ$ and also "slopes" related to these maxima by gradient ascent. The threshold value $k$ is chosen as

$$k = j/J = J/J_M \tag{6}$$

Here, $J_M$ is the maximal value of intensity over the whole field in a moment. In this picture, the brightest maxima are considered separately ($k\sim1$), while weaker data is analyzed as an average over several neighbor maxima. The fragments associated with stronger maxima are separated first. Figure 4 shows the example of the fragmentation procedure corresponding to the image in Figure 1a.

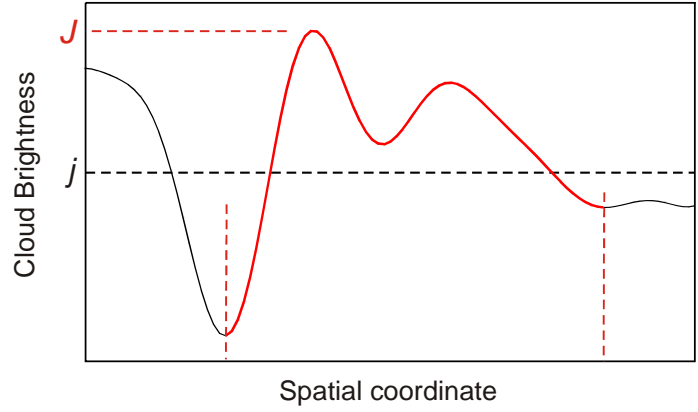

*Figure 3. Graphic scheme of the fragmentation procedure.*

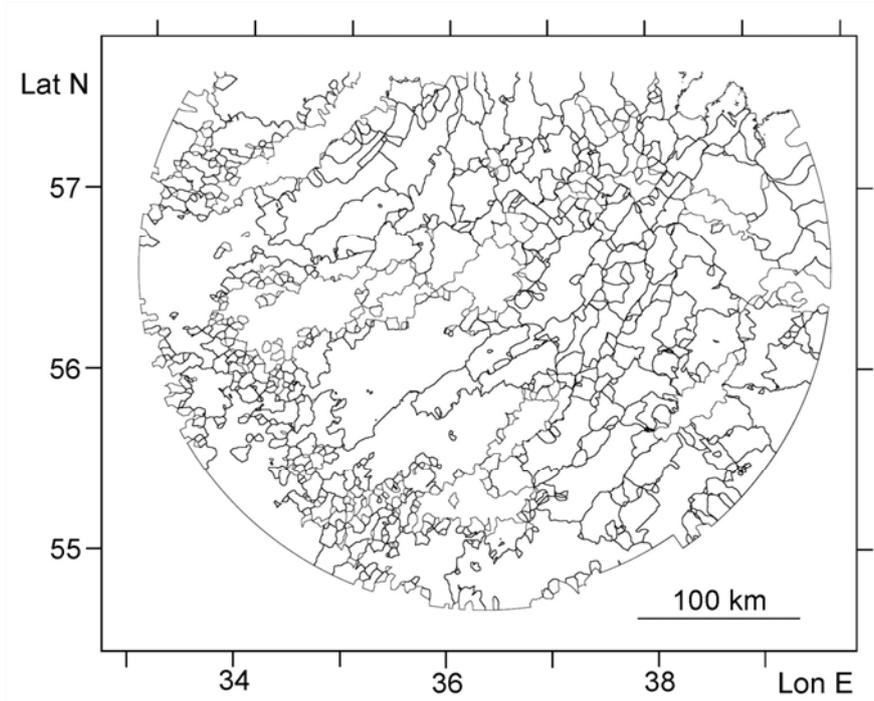

*Figure 4. Example of the result of the fragmentation procedure (the same moment as in Figure 1).*

We denote the Pearson correlation coefficient for functions $F$ and $G$ as $<F(x,y), G(x,y)>$. Taking the pictures from the same camera for the moments $T$ and $T+\Delta T$, we compute the correlation function inside the definite fragment:

$$D(\Delta x, \Delta y, \Delta T) = <J(x, y, T), J(x+\Delta x, y+\Delta y, T+\Delta T)> \qquad (7)$$

It is used to find displacement values of $\Delta x$ and $\Delta y$ corresponding to the maximum $D$ for definite $\Delta T$. Repeating this procedure for all images in the same site in an interval of $\Delta T$ (±1 min), we build the linear relations $\Delta x(\Delta T)$ and $\Delta y(\Delta T)$ and find zonal and meridional velocities of the NLC wave pattern $v_x$ and $v_y$, respectively. The procedure is run independently for both sites, and then average velocities are found. To fix possible errors, i.e., those related to non-horizontal waves and instabilities (see above), the correlation criterion is applied: all correlations (7) during ±1 min interval must be not less than 0.5, and the correlation averaged on this time interval must be not less than 0.75. Actually, for the brightest fragments of NLC field, the correlation usually exceeds 0.9.

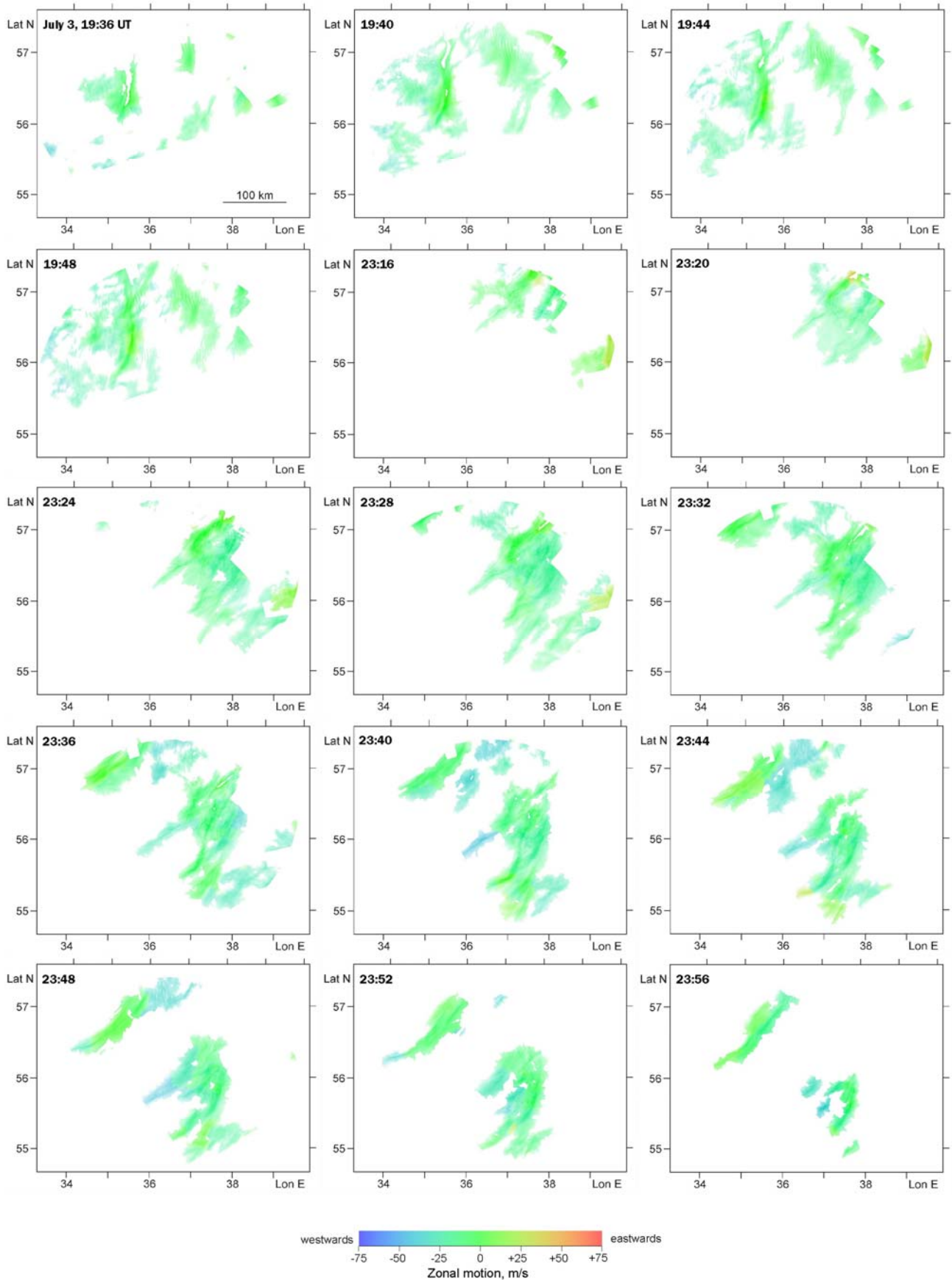

*Figure 5. Maps of cloud pattern zonal velocity during the evening and morning twilight.*

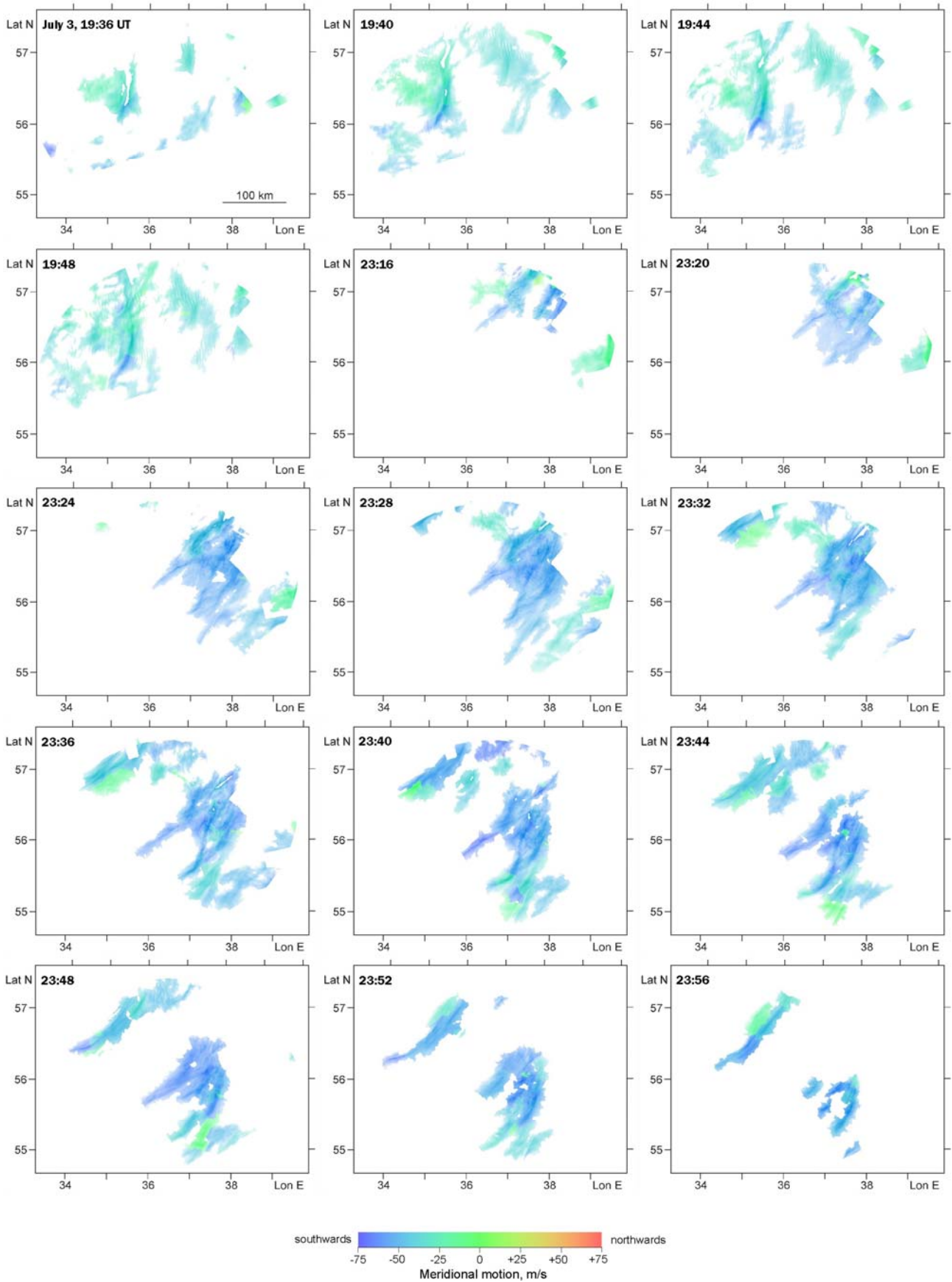

*Figure 6. Maps of cloud pattern meridional velocity during the evening and morning twilight.*

Figures 5 and 6 show the maps of zonal and meridional velocities during the evening and morning twilight. The negative sign of both values indicates the south-west direction of the motion. The same (with a little bit higher value of velocity) was registered for NLC in site B in 2022 (Ugolnikov, 2023b) and during the balloon NLC measurements (Dalin et al., 2022). Southward winds bringing the colder air masses from the north are typical for the case of bright NLC (Fiedler et al., 2011, Gerding et al., 2021).

When velocities are found, we can take the sum of all pictures on the same site in a time interval ±1 min:

$$J_{S(A,B)}(x,y,T) = \frac{1}{N_{A,B}} \sum_{\Delta T} J_{A,B}(x+v_x\Delta T, y+v_y\Delta T, T+\Delta T) \quad (8)$$

Here $N_{A,B}$ is the number of frames in this interval taken at sites A and B. These sums are used to find the triangulation altitude of the NLC fragment. We introduce the coordinate system $(p, q)$, where the $q$-axis is directed along the baseline, and $p=q=0$ in the middle of the baseline (see Figure 1). Taking into account the small altitudes of observation sites, the difference between real and *a priori* altitudes, $\Delta H$, causes the shift of pictures parallel to the observational baseline (by $q$ coordinate in Figure 1). We have to find the shift value $\Delta q = q_A - q_B$ corresponding to the maximal correlation $<J_{SA}, J_{SB}>$.

The principal scheme of altitude definition for the simple case (cloud above the baseline) is shown in Figure 7. Knowing the altitudes of observation sites and using the fact that all values of site altitudes $h_A$ and $h_B$ and $\Delta H$ are significantly less than $H_0$, we refer the observational baseline $L_0$ to the sea level:

$$L = L_0 + \frac{1}{H_0}\left(h_A\left(\frac{L_0}{2}-q\right) + h_B\left(\frac{L_0}{2}+q\right)\right) \quad (9)$$

Since altitudes $h_A$ and $h_B$ are small, the picture remains the same if the NLC fragment is not above the baseline ($p\neq 0$, see Figure 1) and the plane of Figure 7 is not vertical. The approximate "flat" value of the NLC altitude correction is

$$\Delta H_F = \Delta q \frac{H_0}{L - \Delta q} \quad (10)$$

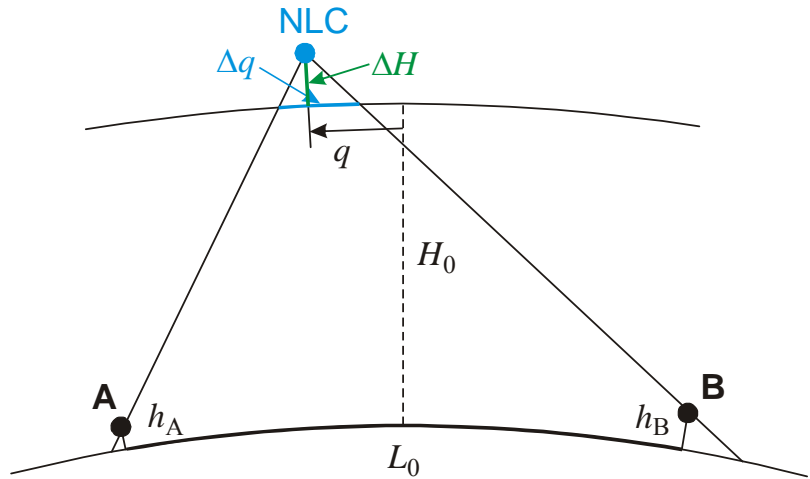

*Figure 7. Scheme of altitude definition for the simple case if the cloud is above the baseline.*

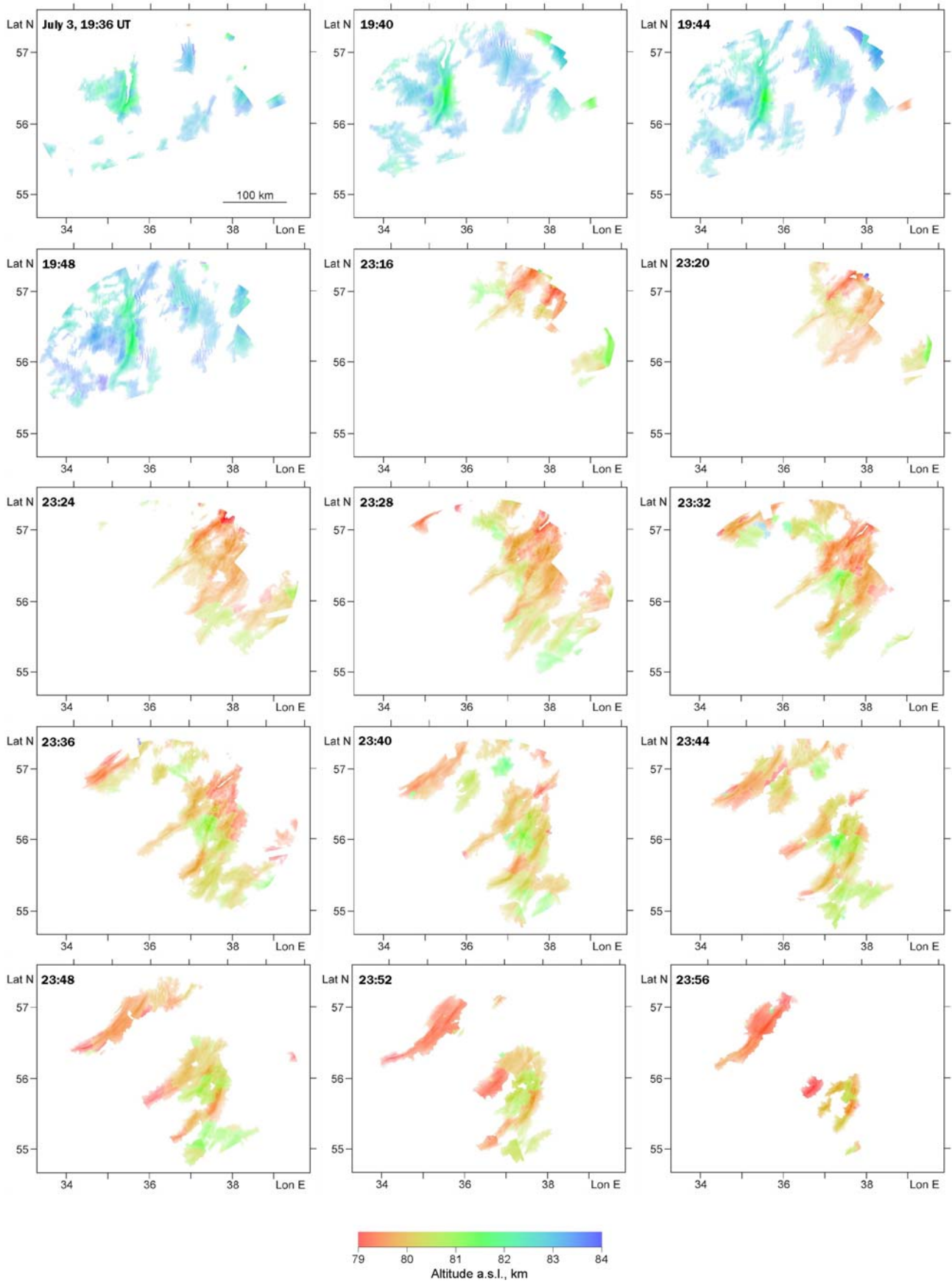

*Figure 8. Maps of cloud altitude during the evening and morning twilight.*

If this value does not exceed several kilometers, the curvature of the Earth can be taken into account in a simplified form:

$$\Delta H = \Delta q \frac{H_0}{L - \Delta q} \cdot \left(1 + \frac{L^2}{8RH_0} + \frac{p^2 + 3q^2}{2RH_0}\right) \quad (11)$$

Here $R$ is the mean Earth radius, $p$ and $q$ are NLC fragment coordinates. For $\Delta H \sim 2$ km non-flat correction of NLC altitude is not more than 0.05-0.06 km. The accuracy of approximation (11) is better than 0.01 km, which is enough for our analysis.

Figure 8 shows the maps of NLC altitudes at the same moments of time as in Figures 5 and 6. We see that triangulation analysis confirms significant NLC downshifting from the evening until the morning that was fixed by color analysis for the same night, the evening-morning difference is about 2.5 km. However, triangulation altitudes are systematically lower than colorimetric altitudes, the reasons will be discussed below. The brightest NLC fragments are typically lower than their surroundings, this is seen during both evening and morning twilight. It is interesting in relation with particle size variations, the procedure of effective radius estimation is described in the following chapter.

**4. Particle size mapping**

Following Equations (3) and (4), we see that if we fix the same NLC fragment from two sites taken in the same time and background noise and wave disturbances do not affect the correlation, then the ratio of normalized intensities $J_{(1,2,3)N}$ (now we drop out the index N) is equal to the ratio of Mie scattering coefficients $S_{1,2,3}$ normalized by the same ratio for *a priori* particle radius taken for normalization, $r_0 = 85$ nm. For the base length equal to 114.7 km, the difference in scattering angles can be up to 70º, scattering ratio significantly depends on particle size. Influence of non-horizontal wave structure on the color of NLC is weaker than on its intensity, and we take one intensity ratio and two color ratios as observational parameters used for size definition:

$$C_1 = \frac{J_{A1N}}{J_{B1N}} = \frac{S_1(\theta_A, r)/S_1(\theta_B, r)}{S_1(\theta_A, r_0)/S_1(\theta_B, r_0)}, \quad C_2 = \frac{J_{A2N}/J_{A1N}}{J_{B2N}/J_{B1N}} = \frac{(S_2/S_1)(\theta_A, r)/(S_2/S_1)(\theta_B, r)}{(S_2/S_1)(\theta_A, r_0)/(S_2/S_1)(\theta_B, r_0)},$$

$$C_3 = \frac{J_{A3N}/J_{A1N}}{J_{B3N}/J_{B1N}} = \frac{(S_3/S_1)(\theta_A, r)/(S_3/S_1)(\theta_B, r)}{(S_3/S_1)(\theta_A, r_0)/(S_3/S_1)(\theta_B, r_0)} \quad (12)$$

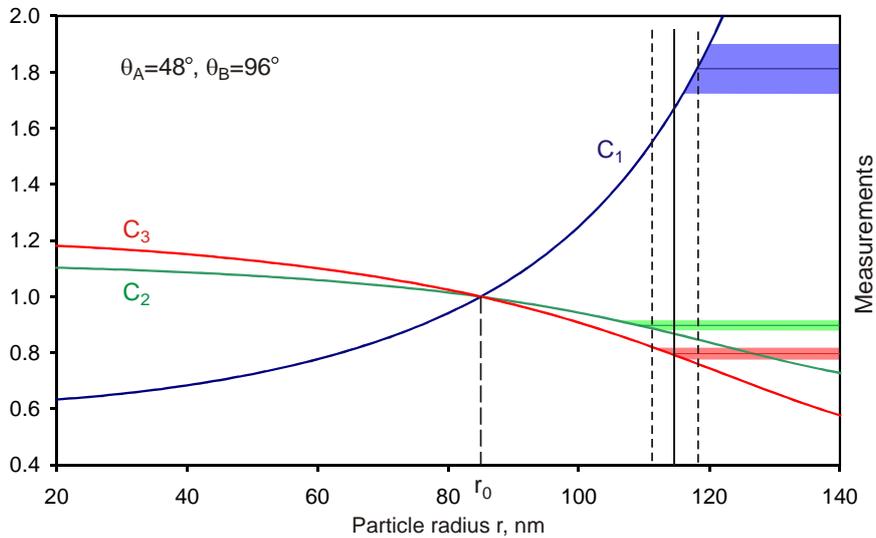

*Figure 9. Example of effective particle radius definition.*

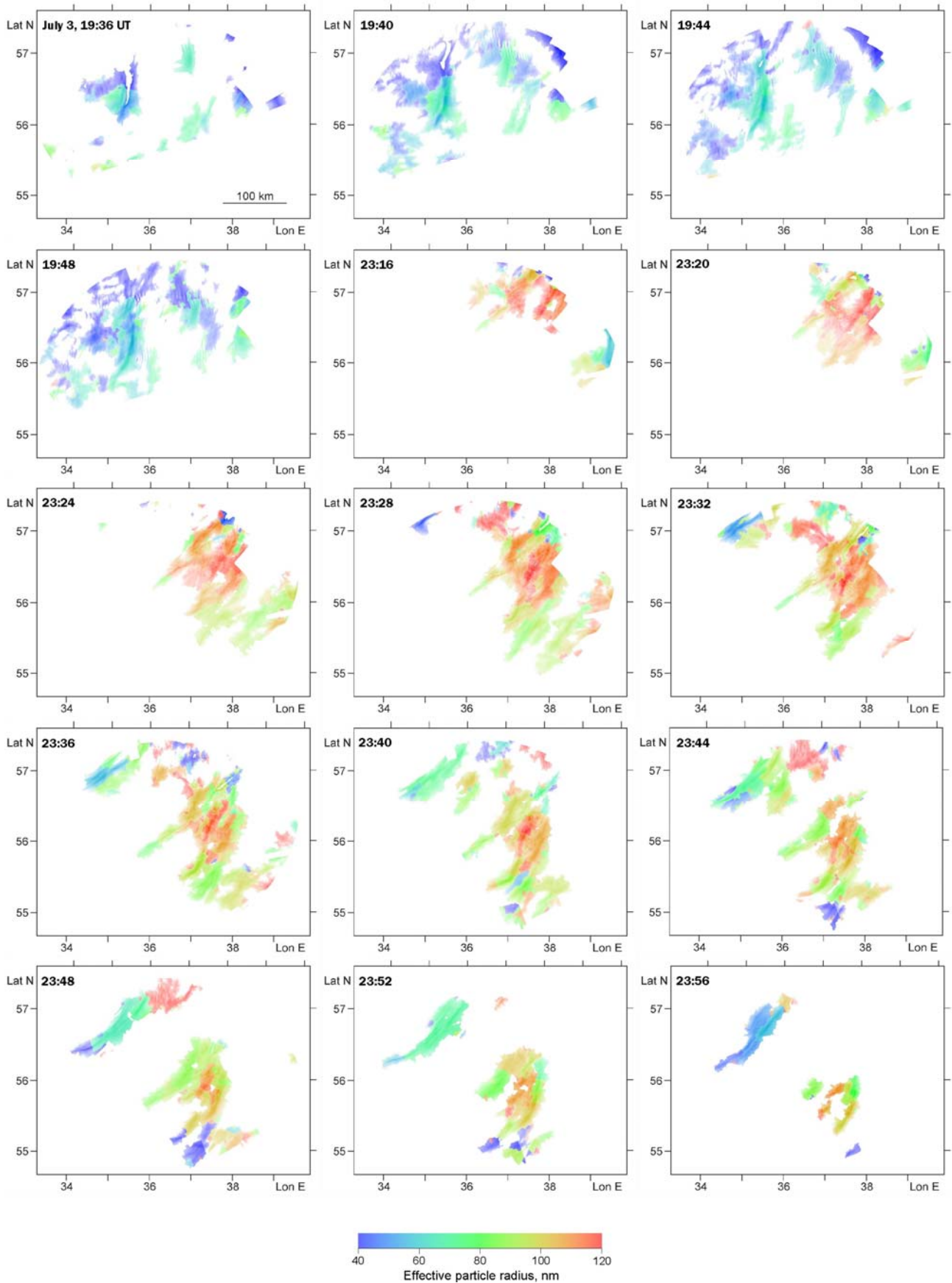

*Figure 10. Maps of effective particle radius during the evening and morning twilight.*

Indexes 1, 2, and 3 refer to B, G, and R channels, respectively. If the effective particle size is equal to $r_0 = 85$ nm, all indexes will be equal to unity. Theoretical values $C_{T1,2,3}$ are calculated by Mie theory for given scattering angles $\theta_{AB}$. If $C_{O1,2,3}$ are measured values and $\sigma_{1,2,3}$ are their errors, we find the effective particle radius $r$ by least squares method:

$$\frac{(C_{O1} - C_{T1}(r))^2}{\sigma_1^2} + \frac{(C_{O2} - C_{T2}(r))^2}{\sigma_2^2} + \frac{(C_{O3} - C_{T3}(r))^2}{\sigma_3^2} = min \quad (13)$$

The graphic scheme of this estimation is shown in Figure 9. Experimental data is averaged inside the NLC fragment separated during the triangulation analysis, the shift $\Delta q$ between the fields fixed in two sites is taken into account. Figure 10 shows the maps of effective particle radius during the evening and morning twilight. We note that if we assume the lognormal distribution of particle radius with width 1.4 (von Savigny and Burrows, 2007), then the median radius will be about 0.5 of the effective radius shown in Figure 10.

The first thing that can be seen is a significant increase of particle size in the morning twilight. The central bright spot of the remarkable wave picture observed in the morning contains particles with an effective radius up to 120 nm. Its altitude is close to 80 km, this is known as the typical lower border of the ice freezing layer in the mesosphere, where particles reach their maximal size (Baumgarten et al., 2010). There are also low (<80 km) fragments containing relatively small sublimating particles below the frost layer, the most remarkable is the cloud on the top left of the morning maps.

The last effect is seen in Figure 11 showing the diagram "particle radius – altitude" for most bright and stable cloud fragments during the evening and morning twilight. Asterisk symbols refer to mean values obtained by colorimetric method (Ugolnikov, 2023a). Altitude offset of about 1 km is seen in the color data, possible reasons can be related to uncertainties in OMPS aerosol data or the radiative transfer picture, it will be considered below. Results also reflect the general effects of the NLC diurnal cycle, which are also discussed in the following chapter of the work.

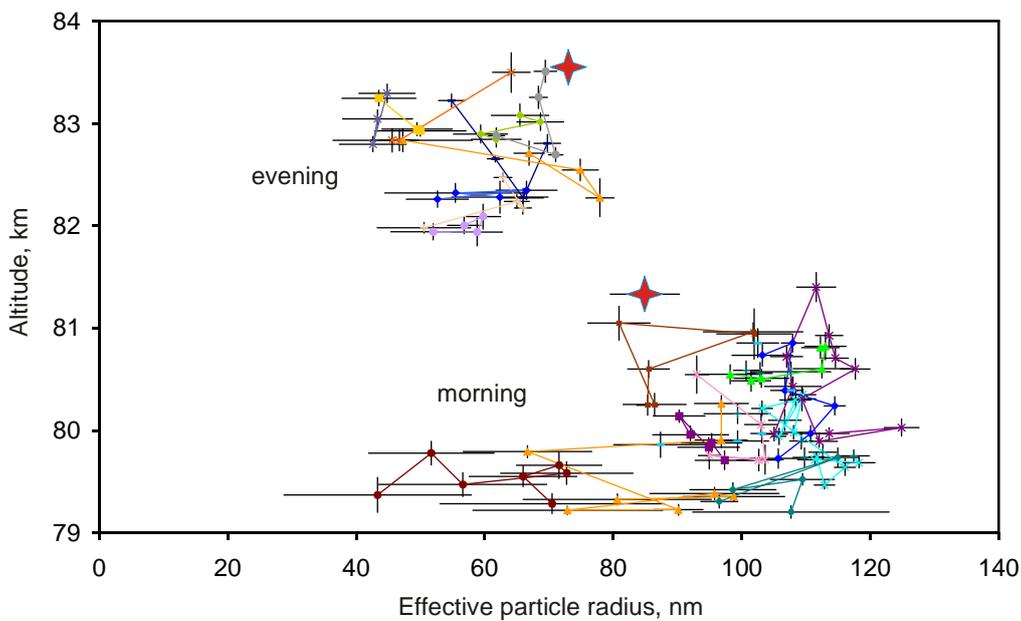

*Figure 11. Evolution of altitude and effective particle radius of stable bright fragments of NLC during the evening and morning twilight. Asterisks show the mean values by color analysis.*

## 5. Discussion

Single-site RGB photometry of noctilucent clouds allowed finding the altitude and particle size averaged over the whole observable cloud ensemble (Ugolnikov et al., 2017; Ugolnikov, 2023a) or along the gravity wave (Ugolnikov, 2023b). Spread RGB photometry by two or more devices with the baseline longer than 100 km makes it possible to do it for each NLC fragment and to build the maps of altitude and particle radius. This was done for the night of July 3, 2023, when noctilucent clouds were the brightest over the entire summer of 2023 in the observation place. It was also the first time at this location during the past several years that bright NLC were successfully observed both in the evening and in the morning twilight.

We can add that the first days of July correspond to the maximal occurrence rate of NLC in the Northern hemisphere (see Robert et al. (2009) or Dalin et al. (2019) for example), and July 3, 2023, was the day of the full Moon. It did not cause the observational problems since the Moon was far below the observable ensemble of NLC in the sky in both sites. The contribution of the moonlight to the sky background is small during the twilight (it is not noticeable in Figure 1, for example), and it was subtracted together with the total background by the procedure described above. However, tidal effects can decrease the temperature and increase the occurrence rate of NLC in these conditions (Kropotkina and Shefov, 1976; Dalin et al., 2006; Pertsev et al., 2015).

The basic effect of the diurnal variability of NLC occurrence rate and brightness is the solar UV-emission absorbed by molecular oxygen and ozone and changing the temperature. This cycle, often called the "solar tidal cycle", is clearly seen in lidar observation statistics (Fiedler et al., 2011; Gerding et al., 2013, 2021; Ridder et al., 2017; Fiedler and Baumgarten, 2018; Schäfer et al., 2020). The mean altitude of NLC decreases after local midnight. Occurrence rate become higher in the same time, particle size also increases. These processes are related to the cooling and downshifting of cold mesosphere layers during the night and are seen in lidar (Fiedler et al., 2011) and in model (Merkel et al., 2009) data on temperature. The largest particles appear near the lower border of the frost layer, where the water vapor density increases. The altitude of bright NLC is systematically lower than the mean value (Gerding et al., 2021).

Downshifting of NLC layer during the night is fixed by triangulation. It is also seen by color analysis based on the observations of the clouds' immersion into and escape from the umbra of the dense atmospheric layers of the Earth (Ugolnikov, 2023a). Downshifting value is the same for both techniques, but 1-km offset between color and triangulation altitude is seen. The possible explanation can be related to the transfer of tangent radiation in the atmosphere. The umbral method was based on the single scattering model. It is known that NLC are unseen in the shadow of the Earth, and observable color changes of clouds are well described by single scattering assumption. However, numerical modeling by Lange et al. (2022) points to a noticeable contribution of multiple scattering at solar zenith angle about 98°, when NLC are emitted by the Sun through the lower stratosphere. If this component basically consists of the scattering of Rayleigh limb emission by NLC particles, it will have a more bluish color and slow reddening of cloud when it immerses into the umbra. This can cause a positive shift of measured umbral altitude; this is the question to be checked in future color and triangulation measurements.

## 6. Conclusion

Spread or net observations of noctilucent clouds by a number of ground-based cameras give the cost-effective opportunity to build the maps of wave pattern velocities, altitude and effective particle radius. While particle size estimation requires accurate background subtraction and field photometry, altitude measurements allow using the photo images from different locations, however, exact optical field conversion parameters must be defined based on nighttime star astrometry.

The choice of baseline is important. Short base lengths (<60-80 km) hardly provide good accuracy of altitude measurements. If the baseline is very long (>150-200 km), the overlapping area of NLC fields observed by two cameras will decrease, and the effects of Earth's curvature (and even non-sphericity) get amplified. The best solution is a net array, with the distance between neighbor observation sites about 80 km. If particle size measurements are planned, the meridional direction of the baseline is effective since it provides the maximal difference of scattering angles for the same cloud fragment.

There is a number of net systems of wide-field cameras installed for NLC observations (Dalin et al., 2008, 2013; Dubietis et al., 2011; Suzuki et al., 2016). An increase in site number and use of high-sensitivity devices with fine angular resolution will make it possible to run the procedure suggested in this work for smaller fragments of cloud fields. This will help to study tiny wave effects and to retrieve 3-d picture of noctilucent clouds.

**Acknowledgments**

Author would like to thank Egor O. Ugolnikov (Silaeder School, Moscow) and Sergey V. Pilipenko (Astro-Space Center of Lebedev's Physical Institute, Moscow) for their help in the measurement process. The work is supported by Russian Science Foundation, grant 23-12-00207.